\documentclass[twocolumn,prb,aps]{revtex4}
\usepackage{amsmath}
\usepackage{amssymb}
\usepackage{graphicx}
\usepackage{dcolumn}
\usepackage{graphicx}
\usepackage{dcolumn}
\usepackage{bm}

\begin{document}

\preprint{Phys.Rev.B}

\title{The resistance of 2D Topological insulator in the absence of the quantized transport}

\author{G.M.Gusev,$^1$ Z.D.Kvon,$^{2,3}$ E.B Olshanetsky,$^2$ A.D.Levin,$^1$ Y. Krupko,$^4$ J.C.
 Portal,$^{4,5}$
 N.N.Mikhailov,$^2$ S.A.Dvoretsky,$^2$}

\affiliation{$^1$Instituto de F\'{\i}sica da Universidade de S\~ao
Paulo, 135960-170, S\~ao Paulo, SP, Brazil}
\affiliation{$^2$Institute of Semiconductor Physics, Novosibirsk
630090, Russia} \affiliation{$^3$Novosibirsk State University,
Novosibirsk, 630090, Russia} \affiliation{$^4$ LNCMI-CNRS, UPR 3228,
BP 166, 38042 Grenoble Cedex 9, France} \affiliation{$^5$ INSA
Toulouse, 31077 Toulouse Cedex 4, France}

\date{\today}
\begin{abstract}
We report unconventional transport properties of HgTe wells with
inverted band structure: the resistance does not show insulating
behavior even when it is of the order of $10^2\times h/2e^{2}$. The
system is expected to be a two-dimensional topological insulator
with a dominant edge state contribution. The results are
inconsistent with theoretical models developed within the framework
of the helical Luttinger liquid.


\end{abstract}

\maketitle

\section{Introduction}
\label{1}

The two-dimensional (2D) topological insulators  (quantum spin Hall
insulator) are characterized by a bulk energy gap and gapless
boundary modes that are robust to impurity scattering and
electron-electron interactions \cite{kane, bernevig, maciejko1}. The
2D quantum spin Hall insulator (QSHI) has been realized in HgTe
quantum wells with inverted band structure \cite{konig, buhmann}.
This novel state results from the intrinsic spin-orbit interaction,
which leads to the formation of the helical edge modes with opposite
spin polarization counter-propagating at a given sample boundary.

In the presence of electro-electron interactions, the edge states of
2D topological insulator (2DTI) can be regarded as an example of
helical Luttinger liquid state (LL) \cite{xu, wu, teo}.

\begin{figure}[ht!]
\includegraphics[width=8cm,clip=]{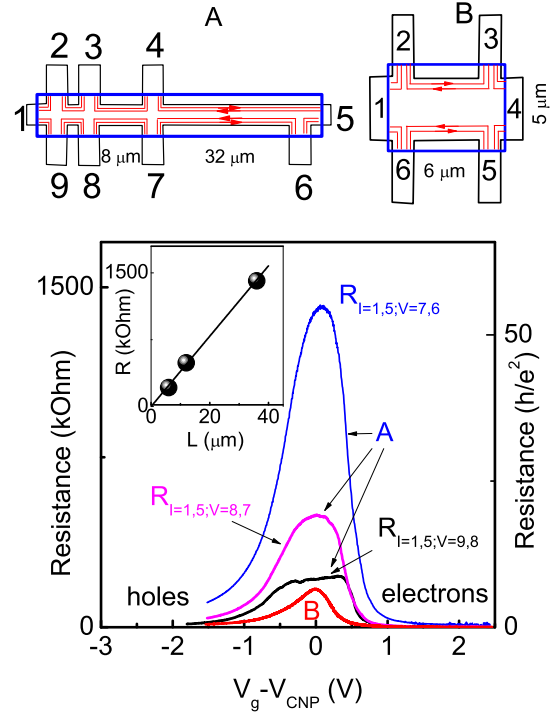}
\caption{\label{fig.1}(Color online) Color online) (a) The
resistance $R_{Local}$   as a function of the gate voltage at zero
magnetic field measured by various voltage probes for samples A and
B, T=4.2 K, $I=10^{-9}$A. The inset shows the the resistance
dependence on the effective distance between probes L. Top
panel-shows a schematic view of the samples. The perimeter of the
gate is shown by the blue rectangle. }
\end{figure}

The helical liquid theory is an extension of the conventional
spinless or spinfull LL theory \cite{voit}. It provides a complete
low-energy description of a one-dimensional metallic system even for
strong interaction and predicts its stability to arbitrary disorder
\cite{xu}. Experimentally the existence of the Luttinger liquid
state in a strongly interacting one dimensional electron system
could be detected from the temperature dependence of its resistance.
The first theoretical model developed within the framework of a pure
LL without impurities in the presence of a single strong barrier
predicted a power law dependence of the resistance on temperature
\cite{kane2}. In a system containing many impurities the resistance
is dominated by a combined effect of interaction and disordered
potential and should exhibit a stronger temperature dependence
\cite{gorny}. The presence of the Luttinger-liquid has been deduced
from the resistance temperature dependence in many mesoscopic 1D
systems, such as single-wall carbon nanotubes \cite{man},
cleaved-edge \cite{auslaender}, split-gate \cite{bell} and V-groove
\cite{levy} semiconductor wires.

The helical edge state in 2D topological insulator is the most
prominent example of an ideal Luttinger liquid, as it is an inherent
property of HgTe wells with an inverted band structure. Furthermore,
it does not require nano manufacturing techniques and its presence
has been observed over the distances of the order of one millimeter
\cite{gusev}. The electronic conductance of the 2DTI is quantized in
units of the universal value $2e^{2}/h$, as was observed in the
short and clean micrometer-scale Hall bars \cite{buhmann}. However,
the quantized ballistic transport has not been observed in sample
with the dimensions above a few microns \cite{konig, buhmann}.
Further investigations are required to understand the absence of the
resistance quantization in macroscopic samples. The evaluation of
the deviation of the conductivity from the quantized value has been
performed in several theoretical models. Particulary, the combined
effect of the weak interaction and disordered scattering in helical
LL channel results to the temperature dependent deviation from
$2e^{2}/h$ which scales with temperature to the power of
 4 \cite{schmidt} or 6 \cite{wu,crepin}. Another way
to explain the observation of 2DTI resistance in excess over its
expected quantized value requires the consideration of possible
scattering processes at the edge. The classical and quantum magnetic
impurity introduces a backscattering between the counter propagating
channels. Such magnetic impurity could be an accidently formed
quantum dot, which can trap odd number of electrons \cite{maciejko}.
For strong enough electron-electron interaction the formation a
Luttinger liquid insulator with a thermally induced transport has
been predicted. Using a somewhat different approach, an edge state
transport theory in the presence of spin orbit Rasba coupling has
developed \cite{strom}. According to this model, the combination of
the spatially nonuniform Rashba spin-orbit and strong
electron-electron interactions leads to localization of the edge
electrons a the low temperatures.

In the present paper we investigate the resistance of the 2D TI with
a dominant edge states contribution to the transport. These edge
channels are described by the helical Luttinger liquid model. The
experiment demonstrates an unexpected weak temperature dependence of
the resistance at a level 100 times higher than the quantum unit
$h/2e^{2}$.

\section{Experiment}
\label{2}

\begin{figure}[ht!]
\includegraphics[width=8cm,clip=]{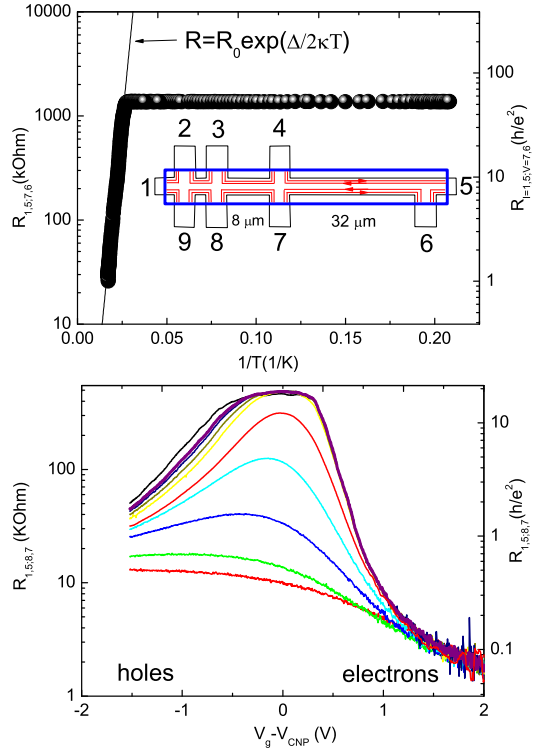}
\caption{\label{fig.2}(Color online) Color online) (a) The
resistance $R_{1,5;7,6}$ corresponding the configuration, when the
current flows between contacts 1 and 5 , and the voltage is measured
between contacts 7 and 6, as a function of the inverse temperature
at charge neutrality point. The solid line is a fit of the data with
the Arrenius function where $\Delta=200$ K. The insert shows a
schematic view of the sample. (b) The resistance $R_{1,5;8,7}$
(I=1,5; V=8,7) as a function of the gate voltage for different
temperatures, (T(K): 1.5, 2.5, 3, 3.5, 4.2, 10, 19, 29, 40, 53, 62),
I=$10^{-9}$A. }
\end{figure}

The $Cd_{0.65}Hg_{0.35}Te/HgTe/Cd_{0.65}Hg_{0.35}Te$ quantum wells
with (013) surface orientations and the width $d$ of 8-8.3 nm were
prepared by molecular beam epitaxy. A detailed description of the
sample structure has been given in \cite{kvon, olshanetsky, gusev2}.
Device A is designed for multiterminal measurements, while device B
is a six-probe Hall bar. The device A consists of three $4 \mu m$
wide consecutive segments of different length ($2, 8 , 32 \mu m$),
and 7 voltage probes. The device B  was fabricated with a
lithographic length $6 \mu m$ and width $5 \mu m$ (Figure 1, top
panel). The ohmic contacts to the two-dimensional gas were formed by
the in-burning of indium. To prepare the gate, a dielectric layer
containing 100 nm $SiO_{2}$ and 200 nm $Si_{3}Ni_{4}$ was first
grown on the structure using the plasmochemical method. Then, the
TiAu gate with the dimensions $62\times8 \mu m^{2}$ (device A) and
$18\times10 \mu m^{2}$ (device B) was deposited. Several devices
with the same configuration have been studied. The density variation
with gate voltage was $1.09\times 10^{15} m^{-2}V^{-1}$. The
transport measurements in the described structures were performed in
an variable temperature insert (VTI) cryostat (the temperature range
1.4-60 K), in  $He_{3}$ cryostat (the temperature range 0.3- 3 K)
and in dilution fridge (the temperature range 0.05- 2 K).  We used a
standard four point circuit with a 3-13 Hz ac current of 0.1-100 nA
through the sample. A typical 100 Mohm resistance connected in
series has been used in order to keep the current constant.

The carrier density in HgTe quantum wells can be varied with the
gate voltage $V_{g}$. The typical dependence of the four-terminal
$R$ resistances of the devices A and B as a function of $V_{g}$ is
shown in Figure 1. The resistance $R$ of device A corresponds the
configuration, when the current flows between contacts 1 and 5 and
voltage is measured between different probes. The measured
resistance exhibits a peak that is much greater than the universal
value $h/2e^{2}$, which is expected for QSHI phase. This value
varies linearly with the distance between probes L (see inset). It
is worth noting that the contacts to QSHI are assumed to be thermal
reservoirs, where the electron states with opposite spins are
mixing. In contrast to the Quantum Hall effect, where the mixing of
the edge states occurs within metallic Ohmic contacts, in our
samples its takes place in the 2D electron gas region outside of the
metallic gate due to a finite bulk conductivity. Therefore the
effective length of the 1D channels L exceeds  the distance between
the probes of the Hall bar by 3-4 $\mu m$. It may be expected that a
reflection occurs when a 1D electron wave hits the interface between
the ungated and the 2DTI regions, which may account for a resistance
greater than $h/2e^{2}$. However, a linear dependence the resistance
on L rules out this possibility and confirms that the high
resistance value very likely is a result of backscattering between
the counter propagating edge channels.

One can see in the figure 1 that the device B shows a smaller and
more narrow resistance peak. The Hall coefficient (not shown)
reverses its sign and $R_{xy}\approx0$ when $R$ approaches its
maximum value \cite{gusev2}, which can be identified as charge
neutrality point (CNP). The variation of the gate voltage changes
the charge carrier content in the quantum wells, driving them from
n-type conductor to p-type conductor via an QHSI state.

\begin{figure}[ht!]
\includegraphics[width=8cm,clip=]{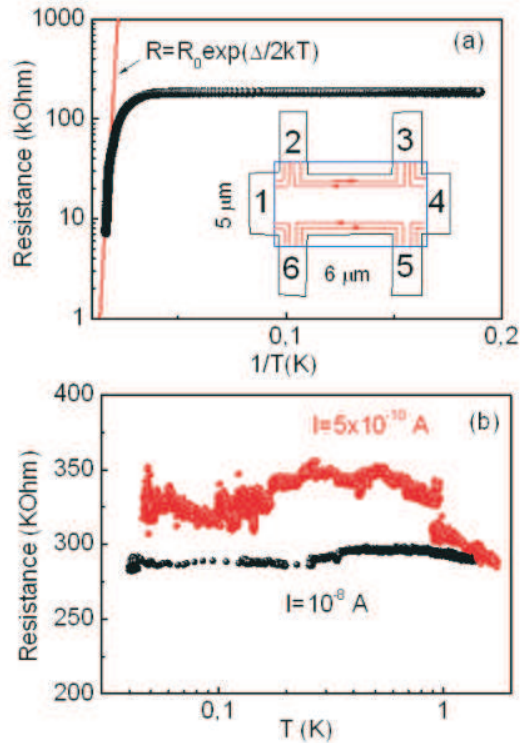}
\caption{\label{fig.3}(Color online) (a) The resistance
$R_{1,4;5,6}$ (I=1,4; V=5,6) of the sample B as a function of the
inverse temperature. The solid line is a fit of the data with the
Arrenius function where $\Delta=400 K$. The insert shows a schematic
view of the sample. (b)  The resistance $R_{1,4;5,6}$ as a function
of the temperature at the charge neutrality point
($V_{g}-V_{CNP}$=0) for
 two different current values.}
\end{figure}

Figure 2a shows the resistance of device A as a function of the
inverse temperature. We see that the resistance sharply decreases
above 25 K while below 20 K it has a tendency towards saturation. We
find that the profile of the $R_{1,5;7,6}$ temperature dependencies
above $T> 25 K$  is very well fitted by the activation law $\sim exp
(\Delta /2kT)$, where $\Delta$ is the activation gap. Figure 2b
shows the evolution of the resistance versus gate voltage dependence
with temperature. We see that the electronic part of the dependence
is weakly dependent on the temperature in accessible temperature
range, while on the hole part the curves shows a strong T-dependence
with a saturation at low T.

The thermally activated behaviour of the resistance above 25 K
corresponds to a gap of 17 meV between conduction and valence bands
in the HgTe well. Recent theoretical calculations based on the
effective $6\times 6$ matrix Hamiltonian predicted the gap $\approx
30 meV$ for 8 nm HgTe wells with a $[013]$ interface \cite{raichev}.
The mobility gap can be smaller than the energy gap due to disorder.
It is worth noting that the disorder parameter, which can reduce the
energy gap in QHSI is rather related to the random deviations of the
HgTe quantum well thickness from its average value \cite{tkachov2},
then to the random potential due to charged impurities. The
saturation of the resistance at low temperature is completely
unexpected, because the electrons are in the supposedly strongly
localized regime, where the electrical resistivity of the system is
two orders of magnitude greater than the resistance quantum
$h/2e^{2}$.

\begin{figure}[ht!]
\includegraphics[width=8cm,clip=]{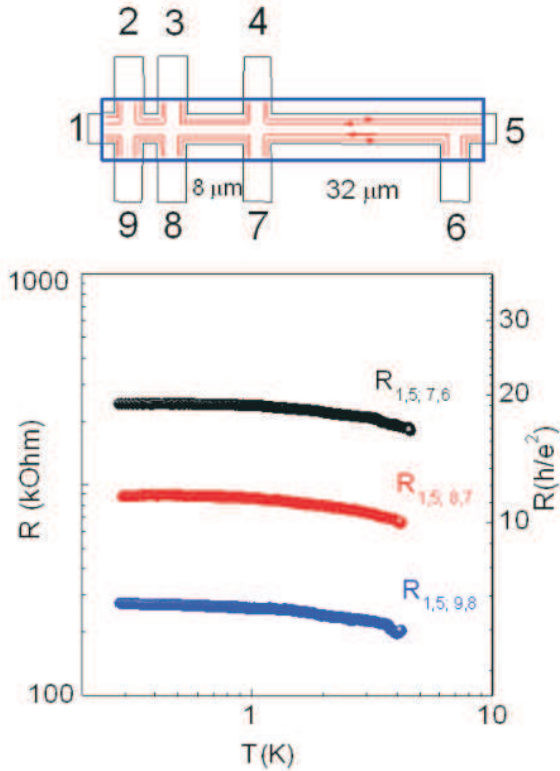}
\caption{\label{fig.4}(Color online) Color online) The resistance
$R$ of the sample A as a function of the temperature at the charge
neutrality point ($V_{g}-V_{CNP}$=0) measured from various voltage
probes in the temperature interval 4-0.3 K, I=$10^{-9} $ A. The top
panel shows a schematic view of the sample.}
\end{figure}

Figure 3a shows the resistance of device B as a function of the
inverse temperature. The data above 25 K is nicely fitted with the
activation dependence, however the activation gap is 2 times larger
than in device A. We attribute the larger value of $\Delta$ to the
better quality of the sample. For example, figure 1 shows that the
resistivity peak in device B is narrower, and it could be argued
that the disorder in this sample is considerably smaller. In order
to prevent  overheating effects by the applied current we study the
current dependence of the resistance. The resistance does not change
very much with the current when the current is varied by three
orders of magnitude and shows saturation at high and low current
values. Such nonmonotonic behaviour is inconsistent with a localized
T-dependence and remains unexplained. In Fig. 3b we present the T-
dependence of $R_{1,4;5,6}$ (I=1,4; V=5,6) at the CNP for two values
of the current measured in a wide temperature range $50 mK < T < 2
K$. We see that at both current levels the resistance is constant
and does not depend on T.

In figure 4 we present the T dependence of the resistance in device
A. We see that the resistance  exhibits the tendency of increasing
with the temperature decreasing, however, it does not show any
significant temperature dependence in the temperature interval (4K
-0.3 K). This behavior is also inconsistent with what is expected
for Anderson localization, Fermi liquid or Luttinger liquid models
\cite{gorny}.

\section{Discussion}
\label{3}

In the rest of the paper we will focus on the several possible
models that can explain the deviation of the resistance from the
expected quantized value. The first one describes the helical edge
liquid state interacting with a single quantum impurity
\cite{maciejko}. The spatially inhomogeneous electrostatic potential
leads to a bound state which traps an odd number of electrons and
forms a magnetic-like impurity. For a large Luttinger parameter
$K>1/4$ corresponding to a weak electron-electron interaction the
conductance is suppressed at low but finite temperatures and is
restored again to a quantized value for $T\rightarrow 0$. For a
strong interaction which corresponds to a small Luttinger parameter
$K<1/4$ the system becomes a LL insulator, and conductance scales
with the temperature as $G(T)\propto T^{2(1/4K-1)}$. Note, however,
that for samples with a top gate parameter K can be estimated from
the expression given in \cite{kane2}. In particular, for our samples
we obtain $K\approx0.6$, which corresponds to a weak coupling
regime.

The second model relies on the localization of electrons due to the
fluctuations of the Rashba spin -orbit interaction caused by charge
inhomogeneity in the presence of the e-e interactions \cite{strom}.
The localization length  strongly depends on the Luttinger parameter
K and can exceed 10 $\mu m$ for $K>0.35$. Note, however, that the
suppression of the conductivity due to localization leads to an
exponential dependence on the temperature. Moreover, Rashba-induced
localization model predicts strong dependence on the sample length
which disagrees with our observations.

In conclusion, we  find that even in an apparently  strongly
localized regime, where the resistance of a HgT quantum well is two
orders of magnitude greater than the resistance quantum $h/2e^{2}$,
the resistance of the 2DTI is independent of temperature indicating
the absence of an insulating phase. The existing theoretical models
do not seem to explain neither such strong deviation from the
quantized value nor the absence of the temperature dependence.

\section{Acknowledgments}
\label{4} We thank M. Feigelman for helpful discussions. A financial
support of this work by FAPESP, CNPq (Brazilian agencies), RFBI
grant N 12-02-00054-a and RAS program "Fundamental researches in
nanotechnology and nanomaterials".

\end{document}